\documentclass[journal]{IEEEtran}

\usepackage{lineno,hyperref}
\usepackage{amsmath,amssymb}
\usepackage{amsthm}
\usepackage{graphicx}
\usepackage{subfigure}
\usepackage{color}
\usepackage{xcolor}
\usepackage{cite}
\usepackage{subfloat}
\usepackage{booktabs}
\usepackage{enumerate}
\usepackage{url}
\usepackage{picinpar}
\usepackage{pifont}
\usepackage{cases} 
\usepackage[font=small,labelfont=bf,labelsep=period]{caption}
\usepackage{epsfig}
\usepackage[latin1]{inputenc}
\usepackage{colortbl}
\usepackage{algorithm}  
\usepackage{algpseudocode}  
\usepackage{amsmath}  

\newcommand{\PreserveBackslash}[1]{\let\temp=\\#1\let\\=\temp}
\newcolumntype{C}[1]{>{\PreserveBackslash\centering}p{#1}}
\newcolumntype{R}[1]{>{\PreserveBackslash\raggedleft}p{#1}}
\newcolumntype{L}[1]{>{\PreserveBackslash\raggedright}p{#1}}

\newcommand{\tabincell}[2]{\begin{tabular}{@{}#1@{}}#2\end{tabular}} 

\newtheorem{Definition}{Definition}
\newtheorem{Remark}{Remark}
\newtheorem{Property}{Property}

\graphicspath{{figure/}}
\begin{document}	
		
\title{Universal chosen-ciphertext attack for a family of image encryption schemes}

\author{Junxin Chen,
	Lei Chen,
	Yicong Zhou,~\IEEEmembership{Senior Member,~IEEE}

	\thanks{This work is funded in part by the National Natural Science Foundation of China (Nos. 61802055, 61771121), by the Macau Science and Technology Development Fund under Grant FDCT/189/2017/A3, and by the Research Committee at University of Macau under Grants MYRG2016-00123-FST and MYRG2018-00136-FST.}	
	\thanks{Junxin Chen is with Sino-Dutch School for Biomedical and Information Engineering, Northeastern University, Shenyang 110004, China (E-mail: chenjx@bmie.neu.edu.cn). He is also with Department of Computer and Information Science, University of Macau, Macau 999078, China.}
	\thanks{Lei Chen is with School of Sciences, Beijing University of Posts and Telecommunications, Beijing 100876, China (Email: clei@bupt.edu.cn).}
	\thanks{Yicong Zhou is with Department of Computer and Information Science, University of Macau, Macau 999078, China (E-mail: yicongzhou@um.edu.mo).}
   }

\markboth{IEEE Transactions on Information Forensics and Security,~Vol.~**, No.~**, **~2019}%
{Shell \MakeLowercase{\textit{et al.}}: Bare Demo of IEEEtran.cls for IEEE Journals}
\maketitle

\begin{abstract}
During the past decades, there is a great popularity employing nonlinear dynamics and permutation-substitution architecture for image encryption.
There are three primary procedures in such encryption schemes, the key schedule module for producing encryption factors, permutation for image scrambling and substitution for pixel modification. 
Under the assumption of chosen-ciphertext attack, we evaluate the security of a class of image ciphers which adopts pixel-level permutation and modular addition for substitution. 
It is mathematically revealed that the mapping from differentials of ciphertexts to those of plaintexts are linear and has nothing to do with the key schedules, permutation techniques and encryption rounds. 
Moreover, a universal chosen-ciphertext attack is proposed and validated. 
Experimental results demonstrate that the plaintexts can be directly reconstructed without any security key or encryption elements. 
Related cryptographic discussions are also given.
\end{abstract}

\begin{IEEEkeywords}
	Cryptanalysis, substitution and permutation, modular addition, chosen-ciphertext attack
\end{IEEEkeywords}

\IEEEpeerreviewmaketitle

\section{Introduction}
Benefiting from the fascinating Internet applications such as twitter, instagram etc., recent years have witnessed a dramatic popularity of the multimedia exchanging over public networks. 
Persons are accustomed to sharing images or videos whenever and wherever possible, which has indeed become an indispensable issue of our daily life. 
On the other hand, this popularity has further led to an increasing requirements of secure transmission and storage of multimedia data over public communication infrastructures.  
Encryption is the easiest way to cope with this issue. 
Obviously, conventional ciphers like Triple-DES and AES can be straightforward employed to encrypt multimedia data by considering it as standard bit stream, this is the so-called `naive encryption' \cite{mmhandbook2004}. 
On the contrary, many researchers strive for the specialized encryption schemes by taking advantage of the intrinsic features of multimedia data \cite{mmhandbook2004,somebasic2006}, such as significant pixel correlation and large data volume.
The encryption schemes analyzed in this paper belong to the latter.
Hereinafter, we focus on image encryption schemes as video ciphers are usually built by operating image ciphers on frame-by-frame fashion.

\begin{figure}
	\centering
	\includegraphics[width=9cm]{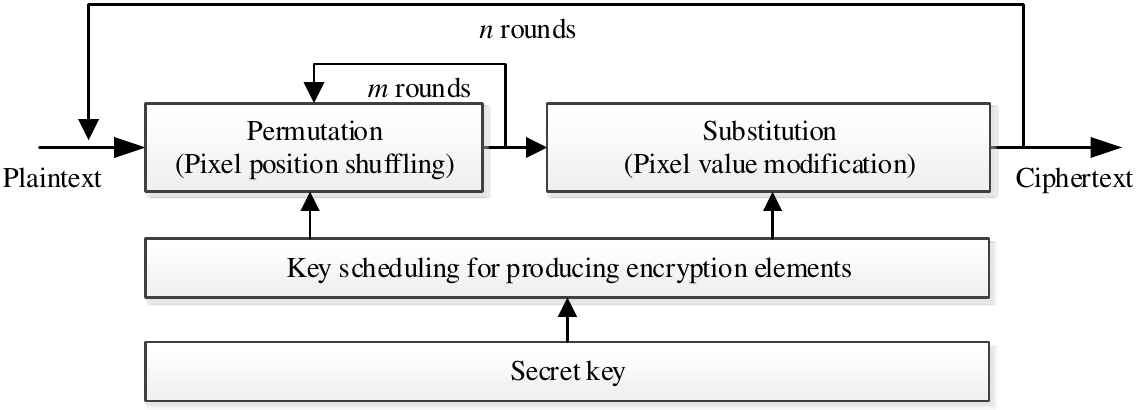}
	\caption{\label{fig:pdstructure} Permutation-substitution architecture for image encryption.}	
\end{figure}

The permutation-substitution \footnote{It may be referred as permutation-diffusion in some literatures, yet the so-called diffusion module is actually a substitution procedure for diffusion.} is the most popular architecture for image ciphers. 
As plotted in Fig. \ref{fig:pdstructure}, an iterative permutation process is implemented for pixel shuffling (with their values unchanged), and then a substitution procedure is performed for modification of pixel values and avalanche performance.
The encryption kernel will repeat many rounds to achieve a higher security level.
Originating from the works of Fridrich and Chen \cite{fridrich1998symmetric,chen2004a3d,chen2004a3dbaker},
chaos and other nonlinear dynamics have been frequently employed to produce the encryption elements required for permutation and substitution.
This is because the fundamental properties of chaotic systems such as ergodicity and sensitivity to initial conditions are desirable for the confusion and diffusion effects of an encryption scheme \cite{somebasic2006}. 
From the past decades, the permutation-substitution structure has aroused a booming of image ciphers.
Their primary innovations can be identified into three categories, i.e., novel permutation approaches \cite{chen2004a3d, chen2004a3dbaker, zhou2013nkp, ZANIN2014288, ARMANDEYEBEFOUDA2014578, Dzwonkowski7185411, praveenkumar2015triple, wu2017anovelaccess, LAN2018133,YicongTC6940279,HUA2018134,lin2015designarm}, new substitution techniques \cite{chen2004a3d, HUA201580, YicongTC6940279, yang2015novel}
and more complex dynamic phenomena
\cite{Kaur8352945, HUA201580, Mannai2015, 7389427Chen, YicongTC6940279, Hua2018IETIE, hua2018cosine, Latif8119911, yang2015novel, wang2014aBrownian, WANG20124033, WU2014122}.

Generally, statistical or empirical tests are introduced for security assessment of these permutation-substitution image ciphers. 
The performance indicators include histogram, adjacent pixel correlation, information entropy \cite{wu2013local}, NPCR and UACI \cite{wu2011npcr}, NIST randomness test, etc. 
In \cite{preishuber2018depreciating}, it is found that these empirical indicators are not powerful conditions for the security declaration.  
Actually, a lot of permutation-substitution image ciphers that passed these tests have been cracked \cite{XIE2017150, solak2010cryptanalysis,chen2015differential, chen2017differential,Li8259437}.
However, relative cryptanalysis achievements are usually case-by-case.
For example, the equivalent permutation key of Fridrich's cipher was retrievable by chosen-ciphertext attack \cite{XIE2017150,solak2010cryptanalysis}, while Li cryptanalyzed an encryption scheme employing the first-order time-delay system \cite{Li8259437}.
On the other hand, generalized cryptanalysis of permutation-only  \cite{li2008a, li2011optimal, jolfaei2016on} and substitution-only \cite{zhang2013cryptanalysis_sbox_only} ciphers was conducted without considering the mutual security promotion of each other.
Few universal cryptanalysis for iterative permutation-substitution image ciphers was reported. 

This paper moves one step further, we present universal cryptanalysis for a family of permutation-substitution image ciphers.
From the outside, they share the following similar features. 
\begin{enumerate}
	\item Both permutation and substitution are included;
	\item the encryption core, i.e., permutation-substitution network, can be iteratively performed with round keys;
	\item image permutation is performed in pixel-level, with independent shuffling matrix;
	\item substitution is implemented with modular addition technique, i.e., Eqs. (\ref{eq:modaddiffusion}) to  (\ref{eq:modaddc2diffusion}) or their variants. 
	\begin{align}
	&c(i)=m(i) \dot + k(i) \label{eq:modaddiffusion}\\
	&c(i)=m(i) \dot + k(i) \dot + c(i-1)  \label{eq:modaddc1diffusion}\\
	&c(i)=m(i) \dot + k(i) \dot + c(i-1) \dot + c(i-2) \label{eq:modaddc2diffusion}
	\end{align}			
\end{enumerate}
Typical ciphers can be found in \cite{LAN2018133,YicongTC6940279,HUA2018134,borujeni2009chaotic,HUA2018148,HUA201580,hua2018cosine,hua2019imageAccess,hua2017filtering}, which all fall into the scope of this work.
Our achievement is different from \cite{preishuber2018depreciating} where Uhl placed more emphasis on the insufficiency of involved statistical/empirical tests for security declaration.
We will theoretically illustrate and experimentally verify that a family of permutation-substitution image ciphers are breakable.
This work can also be considered as the inheritance and development of related works \cite{chen2017differential,zhang2017security,yu2018differential}, because our mathematical analysis and universal practicability are better than both counterparts \footnote{Specifically, the work in \cite{chen2017differential} requires mathematically generalization, while the cryptanalysis in \cite{zhang2017security} should be better extended to more encryption rounds.}. 
Besides, the proposed attack is feasible to recover the plaintext without retrieving the permutation vectors and substitution masks in advance.
This is significantly different from the previous achievements. 

Our contributions are summarized as follows. 
\begin{enumerate}
	\item We formalize the security of a family of permutation-substitution image ciphers which adopts pixel-level permutation and modular-addition substitution.
	\item Cryptanalysis of these ciphers is mathematically conducted, and the security flaw is generalized. 	
	\item A chosen-ciphertext attack is proposed, and its universality is discussed.
	\item We further reveal that the security level of involved ciphers cannot be improved by enhancing complex dynamics, sophisticated permutation techniques or increasing the encryption iterations. 
	\item The cryptanalysis and chosen-ciphertext attack are experimentally validated. 	
\end{enumerate}

The remainder of this paper is organized as follows. 
Section \ref{sec:notation_assumption} gives the employed notations and assumption, while related works and the motivations are discussed in Section \ref{sec:related_work}. 
From a basic model, Section \ref{sec:problemfomu} draws the chosen-ciphertext attack, whose universality is comprehensively analyzed in Section \ref{sec:applicability_ana}.
Cryptographic applications and numerical results are given in Section \ref{sec:applications}, finally, Section \ref{sec:conclusions} concludes the paper. 

\section{Notations and assumptions}
\label{sec:notation_assumption}
\subsection{Notations}
\label{sec:notations}
Unless otherwise indicated, most of the notations adopted in this paper are listed in Table \ref{table:notation}.
\begin{table}[!h]\renewcommand{\arraystretch}{1.2}
	\centering
	\caption{Summary of the adopted notations.}
	\begin{tabular}{||p{0.1\linewidth}|p{0.2\linewidth}|p{0.5\linewidth}||}
		\hline
		Notation & Style & Description \\
		\hline\hline		
		$\textbf{\textit{A}}$ & bold uppercase & an assemble, generally denotes a vector \\ \hline
		$a$ & lowercase & element of corresponding assemble in bold uppercase \\ \hline
		$A$ & capital & a constant \\ \hline 
		$\textbf{\textit{X}}^{(i)}$ & superscript bracket-within-number & factors in $i^{th}$ encryption round \\ \hline
		$\textbf{\textit{X}}_i$ & subscript & index of factors \\ \hline
		$\dot{()}$	& algebraic operators with a dot on the head & modulo algebra as described in the following equations, where $G$ represents gray-scale of the plaintext's pixel.
		\begin{eqnarray}
		& &a \dot + b=(a+b) \bmod G \notag\\
		& &a \dot - b=(a-b) \bmod G \notag \\
		& &a \dot \times b=(a\times b) \bmod G \notag \\
		& & {\dot \sum} _{i = 1}^j {{a(i)}}=a(1) \dot + \dots \dot + a(j) \notag
		\end{eqnarray}		 
		\\ \hline		
	\end{tabular}
	\label{table:notation}
\end{table}

Examples or complementary descriptions are as follows:
\begin{itemize}
	\item Otherwise indicated,  $\textbf{\textit{M}}$ and $\textbf{\textit{C}}$ are reserved for the plaintext and ciphertext in a cipher, $\textbf{\textit{P}}$ and $\textbf{\textit{D}}$ specifically denote the intermediate ciphertexts in a permutation-then-substitution or  substitution-then-permutation cipher, while $\textbf{\textit{W}}$ and $\textbf{\textit{K}}$ represent the permutation vector and substitution matrix, respectively.
	
	\item The image is assumed having size $H\times W$. Yet, this paper prefers the vector representation of an image, i.e., $\textbf{\textit{M}}=\{m(1), m(2),\dots,m(i),\dots,m(L)\}, L=H\times W$.	
	
	\item As to the superscript and subscript, for example, $\textbf {\emph{C}}_{1}^{(1)}$ is the ciphertext of $\textbf {\emph{M}}_{{1}}$ in the first round of encryption, whereas $\textbf {\emph{P}}_{2}^{(3)}/\textbf {\emph{D}}_{2}^{(3)}$ denotes the intermediate permutation/substitution ciphertext of $\textbf {\emph{M}}_{{2}}$ in the $3^{rd}$ iteration.
	
	\item As special cases, $\textbf{\emph{M}}^{(1)}$ means the input plaintext,  and the output ciphertext is represented as $\textbf{\emph{C}}^{(N)}$ where $N$ refers to the iteration counts of an encryption scheme.
	
	\item The modular subtraction of two images is defined as their differential, denoted as $\Delta \textbf{\textit{M}}=\textbf{\textit{M}}_1\dot{-}\textbf{\textit{M}}_2$.
\end{itemize}

\subsection{Attack assumptions}
In a secret communication system, the ciphertext is assumed to be transmitted over public channels. 
Technically, everybody is able to eavesdrop the ciphertext, yet it is unaccessible without the key. 
In this scenario, cryptanalysis refers to the signal retrieval without secret key. 
By the power of the adversary, four kinds of attack models are concluded by Kerckhoffs.
\begin{enumerate}
	\item \textit{Ciphertext-only attack:} the adversary only has a number of ciphertexts.
	\item \textit{Known-plaintext attack:} the adversary has a collection of ciphertexts and their plaintexts.
	\item \textit{Chosen-plaintext attack:} the adversary can construct any plaintexts he want and obtain their ciphertexts. 
	\item \textit{Chosen-ciphertext attack:} the adversary can construct any ciphertexts he want and get their plaintexts.	
\end{enumerate}

Chosen-ciphertext attack is employed in this paper. 
Specifically, arbitrary numbers of ciphertexts and their decryption results are obtainable, whereas we don't own the secret key and cannot directly decrypt the receiving ciphertext either. 
By exploiting the knowledge resides in these chosen-ciphertexts and corresponding plaintexts, the attack is said to be successful if any of the receiving ciphertext can be successfully recovered without the key. 
It should be emphasized that cryptanalysis under other attack models is also important, but it is not the focus of our work.
We construct a universal chosen-ciphertext attack that is valid to a class of image ciphers. 

\section{Related work and motivation}
\label{sec:related_work}

\subsection{Overview of image ciphers}
Since the standardization of permutation-substitution architecture \cite{chen2004a3d,chen2004a3dbaker}, a booming of image ciphers was aroused \cite{ozkaynak2018ND}.
To the best of our knowledge, their innovations can be identified into the following three primary categories.
\begin{enumerate}
	\item \textit{Novel permutation approaches.}
	There are two kinds of permutation techniques.
	The first type treats image pixels as a whole, rather than splitting them into binary stream in advance. 
	Arnold cat map, baker map and standard map are the most popular permutation techniques, they have been employed in the foundation work of Fridrich \cite{fridrich1998symmetric}. 
	To perform pixel-level permutation efficiently, Chen extended the cat map and baker map to three-dimensional \cite{chen2004a3d,chen2004a3dbaker}, while a general gray code has also been exploited for image shuffling owing to its efficient hardware and software implementation \cite{zhou2013nkp,ZANIN2014288}.
	Instead of pixel-by-pixel shuffling, row and column circular permutation \cite{Parvin2016MTA} as well as block shuffling have been developed \cite{ARMANDEYEBEFOUDA2014578}.	
	Other mathematical transforms were also introduced for permutation, for example, quaternion rotation \cite{Dzwonkowski7185411}, Hilbert curve \cite{praveenkumar2015triple} and many others \cite{wu2017anovelaccess,LAN2018133,lin2015designarm}.
	On the other hand, bit-level permutation is implemented by splitting the plain image into a binary matrix.
	In this scenario, bit relocation and pixel value modification effects are simultaneously obtained \cite{zhangweiins}.
	
	\item \textit{New substitution techniques.} 
	Unlike traditional S-box mapping, algebraic operations are always employed for pixel substitution. 
	Typical substitution equations are listed in Eqs. (\ref{eq:modaddiffusion})--(\ref{eq:modaddc2diffusion}) and (\ref{eq:xordiffusion})--(\ref{eq:chendiffusion}), where $\dot +$ and $\oplus$ represents modular addition and bitwise exclusive-OR (XOR), $m(i)$, $k(i)$, $c(i)$ and $c(i-1)$ denotes current plain pixel, secret mask, cipher pixel and the previous cipher pixel, respectively \cite{fridrich1998symmetric}.
	The pixel $c(i-1)$ is introduced for spreading one pixel change in the plaintext to a large region in the ciphertext, i.e., the so-called avalanche effect. 
	\begin{align}
	&c(i)=m(i)\oplus k(i)\oplus c(i-1) \label{eq:xordiffusion}\\
	&c(i)=(m(i)\dot + k(i)) \oplus k(i) \oplus c(i-1) \label{eq:chendiffusion}
	\end{align}	
	Besides the masking formulas, various substitution patterns are also attractive, such as bilateral-substitution \cite{TONG2012850}, simultaneous permutation-substitution \cite{Diab8418363,ENAYATIFAR2017146} and pixel-related avalanche mechanism \cite{Norouzi2014,Chen2018340DNA}.	
	
	\item \textit{Introducing complex dynamic phenomena.} 
	Permutation vector and substitution masks are secret essentials in these ciphers.
	They are generally produced in a secret and random way, with the employed complex dynamics. 
	In the early age, classical chaotic systems such as logistic map and cat map were frequently adopted. 
	Yet, they were assumed insecure in recent years.
	As replacements, hyper-chaotic maps \cite{Kaur8352945} and various improved chaotic principle \cite{HUA201580,Mannai2015,7389427Chen,YicongTC6940279,Hua2018IETIE,hua2018cosine} have been introduced for generating the required encryption elements. 	
	Besides, other complex dynamics originates from physical phenomena also show great potentials.
	For example, quantum walks was evacuated in \cite{Latif8119911,yang2015novel}, whereas Wang introduced Brownian motion \cite{wang2014aBrownian} and wave perturbations \cite{WANG20124033} to produce key stream elements.  
\end{enumerate}

\subsection{Security analysis}
\label{sec:review_cryptanalysis}
Security is the most critical issue for a cipher.
In literature, statistical matrices including histogram, correlation of neighbor pixels, information entropy are usually employed for security evaluation of permutation-substitution image ciphers.
It is very different from the comprehensive mathematical underpinning for traditional ciphers' (such as DES and RSA) security assessment.
The advances of chaotic cryptography reveal that such empirical tests cannot declare security of a cipher \cite{preishuber2018depreciating}, it is easy to construct an apparent insecure encryption scheme and make it pass these tests.
The motivations and security assessment matrices of image ciphers have been depreciated theoretically and experimentally \cite{preishuber2018depreciating}.

In literature, amounts of the `passed' image ciphers have been cracked. 
According to Web of Science, there are approximately 170 publications focus on security analysis of image ciphers \footnote{Search `(attack OR cryptanalysis OR breaking OR cracking OR (security analysis) OR cryptanalyzing OR comment) AND image AND (cipher OR cryptosystem OR encryption)' in the title domain of Web of Science.}. 
Throughout these publications, the following limitations can be identified. 
\begin{enumerate}
	\item \textit{Most of them are case-by-case.}
	In other says, majority of the published attacks are only valid for a specified image cipher, whereas not feasible for counterparts. 
	Fridrich's cipher is cryptanalyzed in \cite{XIE2017150, solak2010cryptanalysis}, which demonstrated a chosen-ciphertext attack for retrieving the permutation matrices. 
	The cipher employing a first-order time-delay system was recently broken by Li \cite{Li8259437} through chosen-plaintext attack,
	and Wang cryptanalyzed two bit-level encryption schemes developed in \cite{chen2015differential, chen2017differential}. 
	Few universal cryptanalysis works were reported.	
	
	\item \textit{Cryptanalysis of permutation-only and substitution-only image ciphers.}
	In \cite{li2008a, li2011optimal, jolfaei2016on}, a permutation cipher has been generalized to an invertible key-dependent vector $\textbf{\textit{W}} = {[w(i) \in \mathbb{L}]}, \mathbb{L}={\{1,2,\dots,L\}}$, whose particle $w(i)$ refers to the secret coordination of a plain pixel that will be re-located to $i^{th}$ position in the ciphertext \footnote{For simplifying the following analysis, $w(i)$ here is essentially defined as the inverse of that in \cite{li2008a,li2011optimal,jolfaei2016on}.}.
	Obviously, $\textbf{\textit{W}}$ is an $\mathbb{L} \to \mathbb{L}$ bijection.
	The function $\cal W$ is further defined to generalize the permutation encryption, as given in Eqs. (\ref{eq:pv_permutation_1D}) and (\ref{eq:W_permutation}).
	\begin{align}
	& p(i)=m(w(i))	\label{eq:pv_permutation_1D}\\
	&\textbf{\textit{P}}=\cal W (\textbf{\textit{M}}) \label{eq:W_permutation}
	\end{align}	
	Known-plaintext and chosen-plaintext attacks have been proved feasible to recover the permutation vector \cite{li2008a, li2011optimal, jolfaei2016on}. 
	Besides, Zhang cryptanalyzed image ciphers which totally use substitution methods \cite{zhang2013cryptanalysis_sbox_only}.
	However, most of the image ciphers include both of the permutation and substitution procedures. 
	The cryptanalysis of permutation-only and substitution-only ciphers were illuminative \cite{li2008a, li2011optimal, jolfaei2016on, zhang2013cryptanalysis_sbox_only}, yet not feasible for the ciphers combining permutation and substitution together. 
	
	\item \textit{Cryptanalysis of a popular substitution mechanism within a single round.}
	As aforementioned, Eqs. (\ref{eq:modaddiffusion})--(\ref{eq:chendiffusion}) are the most popular pixel substitution mechanisms which have been frequently employed in permutation-substitution image ciphers. 
	In \cite{zhang2017security}, Zhang analyzed	the cryptographic strength of Eq. (\ref{eq:chendiffusion}) under the assumption of known-plaintext and chosen-plaintext attacks.
	Subsequently, three image ciphers were cryptanalyzed. 
	However, his cryptanalysis assumed that the encryption was implemented only a single round, and will become invalid for the iterative structure. 
	Except for Zhang's attack \cite{zhang2017security}, most of the cryptanalysis in literature was also achieved by  assuming the encryption as one round. 
	On the contrary, ciphers are generally proposed to be iteratively implemented with round keys, like AES. 	 
\end{enumerate}

\subsection{Motivation and innovation}
Complying with the limitations of the cryptanalysis in literature, we come to the keywords of our work, including `generalized cryptanalysis', `permutation-substitution' and `iteratively performed'. 
They are the primary motivations and innovations of this work. 
In comparison with peer works, the proposed cryptanalysis owns the following features. 
\begin{enumerate}
	\item The cryptanalysis is valid for a class of image ciphers, rather than a specified one.  
	\item Both permutation and substitution procedures are included in the attacked encryption schemes.
	\item The encryption core can be iteratively implemented with round keys. 
\end{enumerate}

Specifically, we focus on the security of a class of image ciphers using pixel-level permutation and modular addition substitution (Eqs. (\ref{eq:modaddiffusion}) to (\ref{eq:modaddc2diffusion})), such as the ones in \cite{LAN2018133,YicongTC6940279,HUA2018134,borujeni2009chaotic,HUA2018148,HUA201580,hua2018cosine,hua2019imageAccess,hua2017filtering}.
We found that they suffer from identical vulnerabilities, nothing to do with the permutation technique, employed nonlinear dynamics, key scheduling process and the encryption rounds. 
A chosen-ciphertext attack is valid to crack all of them without any modification.
Details and discussions will be given in Sections \ref{sec:problemfomu} -- \ref{sec:applications}.

\section{Problem formulation from a basic model}
\label{sec:problemfomu}
We start with the cryptanalysis of a basic cipher, which is constructed according to the focused permutation-substitution structure depicted in Fig. \ref{fig:pdstructure}.
Then we come to its vulnerability against differential cryptanalysis, based on which a chosen-ciphertext attack is step-by-step illustrated.

\subsection{A basic model}
\label{sec:demonstration_basic_cipher}

In this cipher, the plaintext $\textbf{\textit{M}}$ is firstly scrambled using a certain permutation technique, then  the permutation ciphertext is modified according to Eq. (\ref{eq:modaddiffusion}).
The core permutation and substitution operations are iteratively implemented for several rounds to produce the final ciphertext. 
 
Following the nomenclature in Section \ref{sec:notations}, the intermediate permutation ciphertext is obtained as  
$\textbf{\textit{P}}=\cal W (\textbf{\textit{M}})$.
Referring to Eq. (\ref{eq:modaddiffusion}), the ciphertext is calculated according to $\textbf{\textit{C}}=\textbf{\textit{P}}\dot{+}\textbf{\textit{K}},$
where $\textbf{\textit{K}}$ is the substitution masks generated by the employed nonlinear dynamics. 
To be concluded, the ciphertext in a certain encryption iteration is 
\begin{equation}
\label{eq:default_oneround_ciphertext}
\textbf{\textit{C}}=\cal W (\textbf{\textit{M}}) \dot{+}\textbf{\textit{K}}.
\end{equation}
Considering that the encryption process is iteratively implemented, the ciphertext in the $i^{th}$ encryption round can be therefore obtained as
\begin{equation}
\label{eq:i_round_ciphertext}
\textbf{\emph{C}}^{(i)}={\cal W} ^{(i)}(\textbf{\emph{C}}^{(i-1)}) \dot{+}\textbf{\textit{K}}^{(i)},
\end{equation}
where $\textbf{\emph{C}}^{(0)}=\textbf{\textit{M}}^{(1)}$ denotes the input plain image accordingly.

\subsection{Differential analysis}
\label{sec:cryptanalysis_basicmodel}

Suppose that there are two plaintexts $\textbf {\emph{M}}_{1}$, $\textbf {\emph{M}}_{{2}}$  and their ciphertexts  $\textbf {\emph{C}}_{1}$, $\textbf {\emph{C}}_{{2}}$ in a certain encryption round.
Referring to Eq. (\ref{eq:default_oneround_ciphertext}), the differential of the ciphertexts is therefore calculated as
 \begin{equation*}
 {
 	\begin{aligned}
 	{\Delta} \textbf {\emph{C}}&=\textbf {\emph{C}}_{1}\dot{-} \textbf {\emph{C}}_{2}  \\
 	&=({\cal W} (\textbf {\emph{M}}_{1}) \dot{+} \textbf{\textit{K}}) \dot{-} ({\cal W} (\textbf {\emph{M}}_{2})\dot{+} \textbf{\textit{K}} )\\
 	&={\cal W} (\textbf {\emph{M}}_{1}) \dot{-} {\cal W} (\textbf {\emph{M}}_{2})\\
 	\end{aligned}.
 }
 \end{equation*}
 As permutation changes only pixel locations with their values un-modified, and pixels in the same position will be transferred to an identical coordinate in the ciphertexts, therefore 
 $
 {\cal W} (\textbf {\emph{M}}_{1})\dot{-} {\cal W} (\textbf {\emph{M}}_{2})=
 {\cal W} { (\textbf {\emph{M}}_{1}\dot{-}\textbf {\emph{M}}_{2})}={\cal W}({\Delta \textbf {\emph{M}}})
 $
 , so that
 \begin{equation}
 \Delta \textbf {\emph{C}}={\cal W}({\Delta \textbf {\emph{M}}}).
 \label{eq:default1}
 \end{equation}
  
\begin{Definition}
	The differential transfer function (DTF) ${\cal H}(\Delta \textbf{\textit{M}})$ is defined as the mapping from ${\Delta} \textbf {\textit{M}}$ to $\Delta {\textbf {\textit{C}}}$ in a certain encryption round, i.e., 
	\begin{equation*}
	\Delta \textbf {\textit{C}}={\cal H}(\Delta \textbf{\textit{M}}). 
	\label{eq:hndefinition}
	\end{equation*}
\end{Definition}
 
 Referring to Eq. (\ref{eq:default1}), DTF of the basic cipher is 
 \begin{equation}
 \Delta \textbf{\textit{C}}={\cal H}_{(basic)}(\Delta \textbf{\textit{M}})={\cal W}({\Delta \textbf{\textit{M}}}).
 \label{eq:default_transfer}
 \end{equation}
 
 \begin{Property}
 	\label{property:default_bijectivity}
 	$
 	{\cal H}_{(basic)}(\Delta \textbf{\textit{M}})$ has bijectivity, namely $\Delta \textbf{\textit{M}}_1=\Delta \textbf{\textit{M}}_2~\text{if and only if}~{\cal H}_{(basic)}(\Delta \textbf{\textit{M}}_1)={\cal H}_{(basic)}(\Delta \textbf{\textit{M}}_2).
 	$
 \end{Property}
 
 \begin{proof}
 	As well-known, the permutation operation ${\cal W}$ is a one-to-one mapping and thus ${\cal H}_{(basic)}(\Delta \textbf{\textit{M}})={\cal W}({\Delta \textbf{\textit{M}}})$ is a bijection.
 	Hence complete the proof. 
 \end{proof}
 
 \begin{Property}
 	\label{property:default_additivity}
 	${\cal H}_{(basic)}(\Delta \textbf{\textit{M}})$ has modular additivity, that is 
 	\begin{align*}
 	{\cal H}_{(basic)}(\Delta \textbf{\textit{M}}_1)\dot{+}{\cal H}_{(basic)}(\Delta \textbf{\textit{M}}_2)&={\cal H}_{(basic)}(\Delta \textbf{\textit{M}}_1\dot{+}\Delta \textbf{\textit{M}}_2).
 	\end{align*}
 \end{Property}
 
 \begin{proof}
 	Considering that the permutation just contributes pixel re-location, and plain pixels in the same coordinate will be shuffled to an identical position in the ciphertexts, thus
 	\begin{equation*}	
 	\begin{aligned}
 	{\cal H}_{(basic)}(\Delta \textbf{\textit{M}}_1)\dot{+}{\cal H}_{(basic)}(\Delta \textbf{\textit{M}}_2)&={\cal W}({\Delta \textbf{\textit{M}}_1})\dot{+}{\cal W}({\Delta \textbf{\textit{M}}_2})\\
 	&={\cal W}({\Delta \textbf{\textit{M}}_1}\dot{+}{\Delta \textbf{\textit{M}}_2})\\
 	&={\cal H}_{(basic)}(\Delta \textbf{\textit{M}}_1\dot{+} \Delta \textbf{\textit{M}}_2)
 	\end{aligned}.
 	\end{equation*}
 	Proof over.	
 \end{proof}	
 
 \begin{Property}
 	\label{property:default_multiplicative }
 	${\cal H}_{(basic)}(\Delta \textbf{\textit{M}})$ has modular multiplicability, that is 
 	\begin{equation*}
 	{\lambda}\dot{\times}{\cal H}_{(basic)}(\Delta \textbf{\textit{M}})=
 	{\cal H}_{(basic)}(	{\lambda}\dot{\times}\Delta \textbf{\textit{M}}).
 	\label{eq:default_multiplicative }
 	\end{equation*}
 \end{Property}
 
 \begin{proof}
 	Similarly, as only pixel relocation is performed in the permutation phase, 
 	\begin{equation*}	
 	\begin{aligned}
 	{\lambda}\dot{\times}{\cal H}_{(basic)}(\Delta \textbf{\textit{M}})=&{\lambda}\dot{\times}{\cal W}({\Delta \textbf{\textit{M}}})\\
 	=&{\cal W}({\lambda}\dot{\times}{\Delta \textbf{\textit{M}}})\\
 	=&{\cal H}_{(basic)}({\lambda}\dot{\times}\Delta \textbf{\textit{M}})
 	\end{aligned}.
 	\end{equation*}	
 	Proof over.	
 \end{proof}	
 Specifically, ${\cal H}_{(basic)}^{(i)}(\Delta \textbf{\textit{M}})$ denotes the DTF in the $i^{th}$ encryption round. 
 As indicated from Eq. (\ref{eq:i_round_ciphertext}), if different permutation vectors $\textbf{\textit{W}}$ are used, the transfer functions ${\cal H}_{(basic)}^{(i)}(\Delta \textbf{\textit{M}})$ are consequently different from each other.
 However, all of them have bijectivity, modular additivity and multiplicative properties.
 In conclusion, the transfer functions ${\cal H}_{(basic)}^{(i)}(\Delta \textbf{\textit{M}})$ are key-dependent, however,
 their bijectivity, modular additivity and multiplicability properties are key-independent.

 \begin{Definition}
 	The cascaded differential transfer function (CDTF) ${\cal H}^{(1)-(N)}(\Delta \textbf{\textit{M}}^{(1)})$ is defined as the mapping from the differential of input plaintexts, i.e., $\Delta \textbf {\textit{M}}^{(1)}$, to that of output ciphertexts in the $N^{th}$ encryption round, i.e., $\Delta {\textbf {\textit{C}}}^{(N)}$, that is
 	\begin{equation}
 	\Delta \textbf {\textit{C}}^{(N)}={\cal H}^{(1)-(N)}_{(cipher)}(\Delta \textbf{\textit{M}}^{(1)}). 
 	\label{eq:hndefinition1n}
 	\end{equation}
 \end{Definition}
 
 \begin{Property}
 	\label{property:default_transitivity}
 	 ${\cal H}^{(1)-(N)}_{(basic)}(\Delta \textbf{\textit{M}}^{(1)})$ also has bijectivity, modular additivity and multiplicability properties.
 \end{Property}\begin{center}
 
\end{center}
 
 \begin{proof}
 	The proof is given in the Appendix.	
 \end{proof}
 
 \begin{Remark}
 	\label{remark:default}
 	As to the basic cipher, the differential of the output ciphertexts is $\Delta \textbf {\textit{C}}^{(N)}$, it is correlated with the differential of original plaintexts $\Delta \textbf{\textit{M}}^{(1)}$ as  
 	\[\Delta \textbf {\textit{C}}^{(N)}={\cal H}^{(1)-(N)}_{(basic)}(\Delta \textbf{\textit{M}}^{(1)})\]
 	which is a bijective, modular additive and multiplicable function.
 \end{Remark}

\subsection{The proposed chosen-ciphertext attack}
\label{sec:chosen_ciphertext_attack}
As revealed in Remark \ref{remark:default}, the CDTF of the basic cipher has bijectivity, modular additivity and multiplicability properties. 
Benefiting from this, a chosen-ciphertext attack can be launched to retrieve the plaintext. 
The attack consists of five steps whose code implementation can refer to Algorithm \ref{alg:attack}, and described in detail as follows. 

\begin{algorithm}[h]  
	\caption{The proposed chosen-ciphertext attack.}  
	\label{alg:attack}  
	\begin{algorithmic}[1] 
		\Require  
		A ciphertext  $\textbf{\textit{C}}$ 
		\Ensure  
		The plaintext  $\textbf{\textit{M}}$ 
		\State $L=Length(\textbf{\textit{C}})$; 
		\State $\textbf{\textit{C}}_0=zeros(1, L)$; 
		\State $\textbf{\textit{M}}_0=decrypt(\textbf{\textit{C}}_0)$;  
		\For{each $i\in [1,L]$}  
		\State $\textbf{\textit{C}}_i=zeros(1, L)$; 
		\State $c_i(i)=1$; 
		\State $\textbf{\textit{M}}_i=decrypt(\textbf{\textit{C}}_i)$; 
		\State $\Delta \textbf{\textit{M}}_i=\textbf{\textit{M}}_i \dot{-} \textbf{\textit{M}}_0$;
		\EndFor  
		\State $\Delta \textbf{\textit{M}}=zeros(1, L)$; 
		\For{each $i\in [1,L]$}  		
		\State $\Delta \textbf{\textit{M}}=\Delta \textbf{\textit{M}} \dot{+} c(i)\dot{\times}\Delta \textbf{\textit{M}}_i$;
		\EndFor  
		\State ${\textbf{\textit{M}}}=\Delta \textbf{\textit{M}} \dot{+} \textbf{\textit{M}}_0$;
		\\
		\Return $\textbf{\textit{M}}$;   
	\end{algorithmic}  
\end{algorithm}

\begin{enumerate}
	\item Construct $L+1$ chosen-ciphertexts, where $L$ represents the number of pixels in the ciphertext.  
	They are denoted as $\textbf{\textit{C}}_0, \textbf{\textit{C}}_1,\cdots,\textbf{\textit{C}}_i,\cdots,\textbf{\textit{C}}_L$. 
	The ciphertext $\textbf{\textit{C}}_0$ is a zero image, whereas $\textbf{\textit{C}}_i, i\in[1, L]$ is constructed by
	\begin{equation*}
	{{c_i}(j) = \left\{ {\begin{array}{*{20}{l}}
			{1,j = i}  \\
			{0,j \ne i,j \in [1,L]}  \\
			\end{array}}. \right.}  \\
	\end{equation*}
	
	\item Obtain their corresponding plaintexts, represented as $\textbf{\textit{M}}_0, \textbf{\textit{M}}_1,\cdots,\textbf{\textit{M}}_i,\cdots,\textbf{\textit{M}}_L$, respectively. 
	
	\item Calculate the differentials of the plaintexts, i.e., obtain $\Delta \textbf{\textit{M}}_i=\textbf{\textit{M}}_i\dot{-}\textbf{\textit{M}}_0, i\in[1, L].$
	The	plaintext $\textbf{\textit{M}}_0$, together with these differentials $\Delta \textbf{\textit{M}}_1, \cdots \Delta \textbf{\textit{M}}_L$ jointly serve as atoms of the following signal recovery process.
	
	\item For any eavesdropped ciphertext $\textbf{\textit{C}}=\{c(i),i\in[1,L]\}$, assumed its plaintext as ${\textbf{\textit{M}}}$ whose differential between ${\textbf{\textit{M}}_0}$ is  denoted as $\Delta {\textbf{\textit{M}}}$.
	We can obtain
	$\Delta \textbf{\textit{M}}$ according to 
	\[\Delta \textbf{\textit{M}} = {\dot \sum} _{i = 1}^L[c(i)\dot{\times}\Delta \textbf{\textit{M}}_i].\]
	
	\begin{itemize}
		\item It is obviously that 		
		$\textbf{\textit{C}}={\dot \sum} _{i = 1}^L[c(i)\dot{\times} \textbf{\textit{C}}_i]$.
		\item As  $\textbf{\textit{C}}_0$ is a zero image, hence $\Delta \textbf{\textit{C}}_i = \textbf{\textit{C}}_i \dot{-} \textbf{\textit{C}}_0 = \textbf{\textit{C}}_i$.
		\item Further, we can get
		$\Delta \textbf{\textit{C}}={\dot \sum} _{i = 1}^L[c(i)\dot{\times}\Delta \textbf{\textit{C}}_i]$.
		
		\item Referring to the bijectivity, modular additivity and multiplicability of the cipher's CDTF, thus, we can get $\Delta \textbf{\textit{M}}={\dot \sum} _{i = 1}^L[c(i)\dot{\times}\Delta \textbf{\textit{M}}_i]$.
	\end{itemize}
	
	\item Finally, the plaintext $\textbf{\textit{M}}$ is recovered according to 
	\[{\textbf{\textit{M}}}=\Delta \textbf{\textit{M}} \dot{+} \textbf{\textit{M}}_0.\]
\end{enumerate}

\subsection{Example illustration}
\begin{figure*}
	\centering
	\includegraphics[width=15cm]{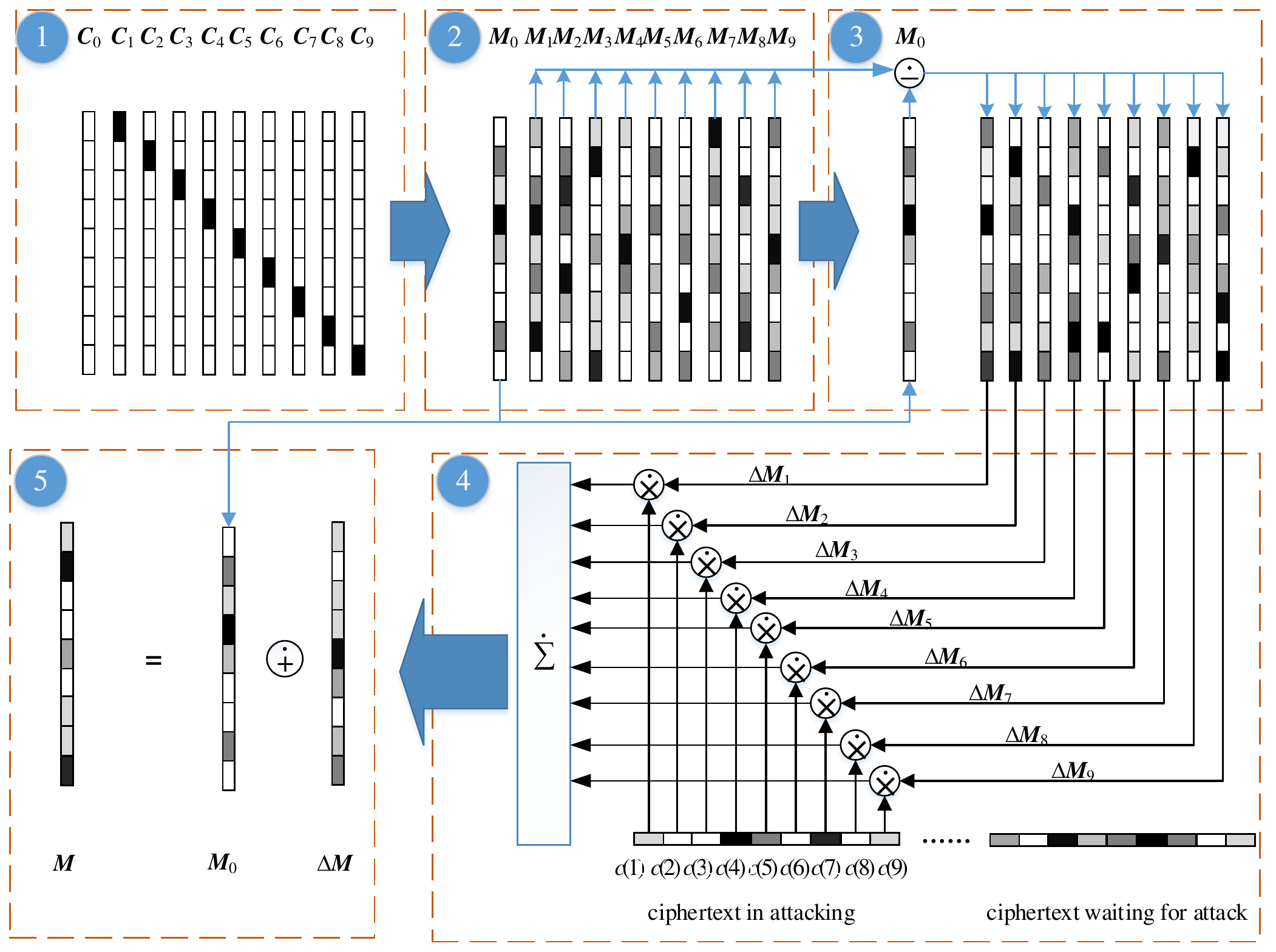}
	\caption{\label{fig:attack} An example illustration of the proposed chosen-ciphertext attack.}	
\end{figure*}

For better understanding, the attack procedures are depicted as Fig. \ref{fig:attack} assuming that the encryption block contains 9 pixels.
Each step of the PCCA is sequentially tagged within a dotted box in Fig. \ref{fig:attack}.

Firstly, $9+1=10$ chosen-ciphertexts are elaborately constructed in the first step, then obtain their corresponding plaintexts in Step 2. 
Unlike the distinctive chosen-ciphertexts, the plaintexts are unpredictable, as revealed by their gray levels. 
In the $3^{rd}$ dotted box, the plaintexts are individually modular subtracted with $\textbf{\textit{M}}_0$.
So that, $\Delta \textbf{\textit{M}}_1, \cdots \Delta \textbf{\textit{M}}_9$ are produced. 
Together with $\textbf{\textit{M}}_0$, they jointly act as atoms of the following signal recovery process.

For a receiving ciphertext $\textbf{\textit{C}}=\{c(1), \cdots, c(9)\}$, modular multiplications are performed between the cipher pixels and corresponding atoms, i.e., $c(i) \dot{\times} \Delta \textbf{\textit{M}}_i$.
Their products are then modular added together to get $\Delta \textbf{\textit{M}}$, retrieved as the differential between the plaintext and $\textbf{\textit{M}}_0$.
This is Step 4 and illustrated in the $4^{th}$ dotted box.
Finally, a modular addition is required to get the recovered plaintext. 

It should be emphasized that once the atoms are produced, any of the receiving ciphertext can be straightforward attacked without repeating the above three processes.
In other words, there may be a waiting list of ciphertexts which are one-by-one recovered according to Steps 4) and 5).

\section{Universality analysis}
\label{sec:applicability_ana}
Hereinafter, we abbreviate the bijectivity, modular additivity and multiplicability as BAM properties, while PCCA is used to denote the proposed chosen-ciphertext attack achieved in Section \ref{sec:chosen_ciphertext_attack}. 

\subsection{Complexity}
Referring to Steps 1) -- 3), PCCA needs $L+1$ chosen-ciphertexts to construct the atoms, i.e., $\textbf{\textit{M}}_0$ and $\Delta \textbf{\textit{M}}_i$.
Once the atoms being established, any of the receiving ciphertext can be straightforward cracked.
In each recovery process, $L$ modular addition and multiplication operations are required to calculate $\Delta \textbf{\textit{M}}$, while another modular addition is sufficient for recovering the plaintext. 

To be concluded, the spatial and computational complexity of PCCA are both $O(L)$.
It should be emphasized that the cost is independent from the encryption rounds $N$.
This is counter-intuitive in comparison with current cryptanalysis achievements whose complexity dramatically increase with the encryption rounds, such as \cite{solak2010cryptanalysis}.

\subsection{Universality}
\label{sec:Universality}
As indicated from Section \ref{sec:chosen_ciphertext_attack}, the CDTF's BAM properties essentially render the feasibility of PCCA.
It is further revealed from  Property \ref{property:default_transitivity} that if a cipher's DTF has BAM features, so is the CDTF.
In conclusion, if a cipher's DTF or CDTF has BAM properties, it is vulnerable against PCCA. 

We find that a family of image ciphers suffer from the aforementioned vulnerabilities, PCCA is hence feasible to crack them directly. 
Their intrinsic DTFs or CDTFs have the BAM properties, while they usually share the following fashions from the outside. 

\begin{enumerate}
	\item The permutation-substitution architecture is adopted, as depicted in Fig. \ref{fig:pdstructure}.
	\item Permutation is performed in pixel-level, whereas modular addition family equations, i.e., (\ref{eq:modaddiffusion}) to  (\ref{eq:modaddc2diffusion}) or their variants, are employed for substitution. 
	\item The permutation vector and substitution mask are key-dependent.
	\item The encryption core can be iteratively implemented with round keys. 
\end{enumerate}

In literature, there are many ciphers of this kind, such as the ones in \cite{LAN2018133,YicongTC6940279,HUA2018134,borujeni2009chaotic,HUA2018148,HUA201580,hua2018cosine,hua2019imageAccess,hua2017filtering} which all fall into the scope of this work.
Besides, the proposed attack is also feasible for permutation-only and substitution-only image ciphers accordingly \footnote{It is different from the specified cryptanalysis given in \cite{li2008a, li2011optimal, jolfaei2016on, zhang2013cryptanalysis_sbox_only}, yet it does work.}.

\subsection{Indications}
Originating from the intrinsic BAM drawbacks of corresponding image ciphers, it is demonstrated that one can recover the plaintext by PCCA, without the permutation vector $\textbf{\textit{W}}$, substitution mask $\textbf{\textit{K}}$ or the launched secret key.
Furthermore, the involved ciphers will also be vulnerable to PCCA no matter how many rounds the encryption are iterated, and no matter the launched keys in each round are different or not. 

This work also indicates that the enhancements in terms of nonlinear dynamics, permutation techniques and substitution masks are useless for the security promotion.
Generally speaking, the only role of the employed nonlinear dynamics (such as the chaotic systems) is to produce the permutation vector $\textbf{\textit{W}}$ and substitution mask  $\textbf{\textit{K}}$.
However, they have been revealed unnecessary in the attack.

In \cite{HUA2018134,hua2017filtering}, a series of random pixels are added around the plaintext before the permutation-substitution network. 
The insertion of random pixels gives indistinguishability of this cipher, i.e., different ciphertexts will be produced at distinct time even with identical plaintexts and secret key. 
In this scenario, known-plaintext and chosen-plaintext attacks become infeasible accordingly.
This is because one cannot control the produced random pixels, hence the actual input of the permutation-substitution iteration is unknown either.
However, such a design is also infeasible to resist against PCCA.
The cipher proposed in \cite{HUA2018134} will be taken as case study and cryptanalyzed in Section \ref{sec:huacryptanalysis}.
To the best of our knowledge, this is the first time to crack an image encryption scheme with noise embedding. 

\section{Cryptographic applications}
\label{sec:applications}
As aforementioned, PCCA is feasible to crack a class of image ciphers. 
Table \ref{table:applicability} lists the brief security analysis of these encryption schemes \cite{LAN2018133,YicongTC6940279,HUA2018134,borujeni2009chaotic,HUA2018148,HUA201580,hua2018cosine,hua2019imageAccess,hua2017filtering}.
All of them adopt pixel-level permutation and employ modular addition for substitution, as indicated in Section \ref{sec:Universality}.
Most of the improvements lie in novel chaotic maps and new permutation techniques, however, they have been indicated unnecessary for the attack.

In the following, two ciphers \cite{YicongTC6940279,HUA2018134} are taken as case studies in individual subsections, while other counterparts will be sketched in Section \ref{sec:cryptan_peers}. 
As the key schedule procedures and permutation techniques have been revealed nothing to do with the resistance against PCCA, they are browsed in the following for simplicity.

\begin{table*}[!h]\renewcommand{\arraystretch}{1.1}
	\centering
	\caption{Applicability of the proposed chosen-ciphertext attack.}
	\begin{tabular}{p{0.09\linewidth}|p{0.36\linewidth}|p{0.3\linewidth}|p{0.07\linewidth}}
		\hline		
		Image ciphers &Encryption process & Security enhancement &  Applicable \\ \hline
		\cite{LAN2018133} & \tabincell{l}{1) substitution with $k(i)\dot{-} m(i)$\\2) key-dependent permutation\\3) four iterations} & \tabincell{l}{1) a new integrated chaotic map \\2) a permutation approach}  & yes \\ \hline		
		
		\cite{YicongTC6940279} &\tabincell{l}{1) substitution using Eq. (\ref{eq:modaddc2diffusion})\\ 2) key-dependent permutation\\ 3) two iterations} &\tabincell{l}{1) a new cascade chaotic map \\2) a permutation approach} & yes \\ \hline		
		
		\cite{HUA2018134} &\tabincell{l}{ 1) insert random pixels around the plaintext's four edges\\ 2) key-dependent permutation\\ 3) substitution using Eq. (\ref{eq:modaddc1diffusion})\\ 4) two iterations} &\tabincell{l}{ 1) inserting random pixels before permutation-\\substitution network \\2) a fast permutation method} & yes \\ \hline
		
		\cite{borujeni2009chaotic} & \tabincell{l} {1) key-dependent permutation\\  2) substitution using Eq. (\ref{eq:modaddiffusion})} & a new permutation method & yes \\ \hline
		
		\cite{HUA2018148} & \tabincell{l}{1) key-dependent permutation \\ 2) substitution using Eq. (\ref{eq:modaddc2diffusion}) \\ 3) four iterations} & \tabincell{l}{1) a new coupling chaotic map \\ 2) a new permutation method} & yes \\ \hline
		
		\cite{HUA201580} & \tabincell{l}{1) key-dependent permutation \\2) row-by-row and then column-by-column  substitution \\using Eq. (\ref{eq:modaddc1diffusion}) \\ 3) two iterations} & \tabincell{l}{1) a new coupling chaotic map \\ 2) a new permutation method \\3) two stages of substitution} & yes \\ \hline
		
		\cite{hua2018cosine} & \tabincell{l}{1) key-dependent permutation \\2) rotation \\3) substitution in random order using Eq. (\ref{eq:modaddc1diffusion}) \\ 4) four iterations} & \tabincell{l}{1) a new coupling chaotic map \\ 2) a new permutation method \\3) the substitution is in random order} & yes \\ \hline
		
		\cite{hua2019imageAccess} & \tabincell{l}{1) key-dependent permutation \\2) substitution using image filtering, i.e., linking many \\neighbor pixels \\ 3) two iterations} & \tabincell{l}{1) a new permutation method \\2) substitution using image filtering} & yes \\ \hline
		
		\cite{hua2017filtering} & \tabincell{l}{1) insert random pixels around two edges \\2) key-dependent permutation \\3) substitution using image filtering \\ 3) four iterations} & \tabincell{l}{1) inserting random pixels before permutation-\\substitution network \\2) substitution using image filtering} & yes \\ \hline
	\end{tabular}
	\label{table:applicability}
\end{table*}

\subsection{Applicability to Zhou's cipher \cite{YicongTC6940279}}
\label{sec:zhoucryptanalysis}
Zhou's ciphr proposed in \cite{YicongTC6940279} that used Eq.~(\ref{eq:modaddc2diffusion}) for substitution is firstly taken as case study.
The encryption processes are given as follows.
\begin{enumerate}
	\item \textit{Initialization}.
	With the key $\textit{Seed}$ and cascade chaotic systems to generate a permutation vector $\textbf{\textit{W}}$ and a substitution mask matrix $\textbf{\textit{K}}$.
	
	\item \textit{Substitution}.
	Two previous cipher pixels are linked for avalanche effect,  that is 
	\begin{equation}    	
	d(i)=\\
	\begin{small} 	
	\left\{{\begin{aligned}{}
		&{m(i)\dot{+}k(i)\dot{+}m(L)\dot{+}m(L-1)} &&i=1\\
		&{m(i)\dot{+}k(i)\dot{+}d(i-1)\dot{+}m(L)} &&i=2\\   
		&{m(i)\dot{+}k(i)\dot{+}d(i-1)\dot{+}d(i-2)}&&i\in [3,L] \\
		\end{aligned}} \right.
	\end{small}.
	\label{eq:Zhoudiffusion}
	\end{equation}	
	\item \textit{Permutation}.
	Shuffle the substitution ciphertext $\textbf{\textit{D}}$ with $\textbf{\textit{W}}$ and a cycle permutation technique. 
	Similarly, the permutation is finalized as Eq. (\ref{eq:Zhoupermutation}).
	\begin{equation}
	\textbf{\textit{C}}=\cal W (\textbf{\textit{D}}) \label{eq:Zhoupermutation}				
	\end{equation}	
	\item \textit{Iteration}. 
	The above procedures are repeated 2 times with different $\textbf {\emph{W}}$ and $\textbf {\emph{K}}$ in each round.
\end{enumerate}

Assume that there are two plaintexts $\textbf {\emph{M}}_{1}$ and $\textbf {\emph{M}}_{{2}}$, their intermediate substitution results $\textbf {\emph{D}}_{1}$ and $\textbf {\emph{D}}_{{2}}$, and the ciphertexts $\textbf {\emph{C}}_{1}$ and $\textbf {\emph{C}}_{{2}}$ in a certain encryption round.
It is obvious that
\begin{equation}
\Delta \textbf {\emph{C}}=\textbf {\emph{C}}_{1}\dot{-} \textbf {\emph{C}}_{2}={\cal W}(\textbf {\emph{D}}_{1}) \dot{-} {\cal W}(\textbf {\emph{D}}_{2})={\cal W}(\textbf {\emph{D}}_{1}\dot{-} \textbf {\emph{D}}_{2})={\cal W}({\Delta \textbf {\emph{D}}}).
\label{eq:zhoudelc1}
\end{equation}
It is not difficult to rewrite the substitution (Eq. (\ref{eq:Zhoudiffusion})) as
\begin{align*}
d(i)=&{\dot \sum} _{j = 1}^i {Fib(i-j+1){\dot{\times}}[m(j)\dot{+}k(j)]}\\
&\dot{+}Fib(i+1){\dot{\times}}m(L)\dot{+}Fib(i){\dot{\times}}m(L-1)
\end{align*}
where $Fib(i)$ means the $i^{th}$ particle of a Fibonacci sequence. 
Therefore, we can get
\begin{equation}
\begin{aligned}
{\Delta}{\emph{d}}(i)=&{\dot \sum} _{j = 1}^i {Fib(i-j+1){\dot{\times}}{\Delta}m(j)}\\
&\dot{+}Fib(i+1){\dot{\times}}{\Delta}m(L)\dot{+}Fib(i){\dot{\times}}{\Delta}m(L-1)
\end{aligned}.
\label{eq:zhoudelta_d}
\end{equation}
Combining Eqs. (\ref{eq:zhoudelc1}) and (\ref{eq:zhoudelta_d}), the DTF of Zhou's cipher can be illustrated as 
\begin{equation}
\left\{{\begin{aligned}{}
&{\Delta \textbf{\textit{C}}= {\cal H}_{(Zhou)}(\Delta \textbf{\textit{M}})= {\cal W}(\Delta \textbf{\textit{D}}) }\\
&\begin{aligned}
{\Delta d}(i)=&{\dot \sum} _{j = 1}^i {Fib(i-j+1){\dot{\times}}{\Delta m}(j)}\\
&\dot{+}Fib(i+1){\dot{\times}}{\Delta m}(L)\dot{+}Fib(i){\dot{\times}}{\Delta m}(L-1)
\end{aligned}
\end{aligned}} \right.
.\label{eq:zhou_transfer}
\end{equation}

It is not difficult to conclude the BAM properties of ${\cal H}_{(Zhou)}(\Delta \textbf{\textit{M}})$. 
\begin{enumerate}
	\item \textit{Bijectivity}. 
	Referring to Eq. (\ref{eq:zhou_transfer}), it is obvious that the mapping from ${\Delta \textbf{\textit{M}}}$ to $\Delta \textbf{\textit{D}}$ is revisable, while ${\cal W}$ also gives a bijection from  ${\Delta \textbf{\textit{D}}}$ to $\Delta \textbf{\textit{C}}$ in Eq. (\ref{eq:zhoudelc1}). 
	Therefore, the mapping between ${\Delta \textbf{\textit{C}}}$ and $\Delta \textbf{\textit{M}}$, i.e., $\Delta \textbf{\textit{C}}={\cal H}_{(Zhou)}(\Delta \textbf{\textit{M}})$, is bijective.
	
	\item \textit{Modular additivity and multiplicability}. 
	As can be observed, ${\cal H}_{(Zhou)}(\Delta \textbf{\textit{M}})$ is a combination of permutation, modular addition and multiplication operations.
	All of them are apparently modular additive and multiplicable, as a consequence, ${\cal H}_{(Zhou)}(\Delta \textbf{\textit{M}})$ has modular additivity and multiplicability.
\end{enumerate}

To be concluded, the DTF of Zhou's cipher \cite{YicongTC6940279} has BAM properties and thus vulnerable against PCCA.

\subsection{Applicability to Hua's cipher \cite{HUA2018134}}
\label{sec:huacryptanalysis}
In \cite{HUA2018134}, an image cipher using Eq. (\ref{eq:modaddc1diffusion}) for substitution is proposed. 
Unlike most of the counterparts, random pixels are firstly generated and pasted around the plaintext, and then the enlarged image is encrypted by the permutation-substitution network \footnote{Two ciphers are proposed in \cite{HUA2018134}, this paper focus MIE-MA which uses modular addition for pixel substitution.}. 
The insertion of random pixels gives indistinguishability of the cipher, so as to resist against known-plaintext and chosen-plaintext attacks.
After cryptanalyzing Zhou's cipher that used Eq. (\ref{eq:modaddc2diffusion}) for substitution, it is reasonable to infer that similar encryption scheme adopting substitution with Eq. (\ref{eq:modaddc1diffusion}) is also vulnerable against PCCA.
To some extent, Hua's cipher \cite{HUA2018134} is introduced with specific motivation to show that such a random inserting process is infeasible to resist against PCCA.
The encryption processes in \cite{HUA2018134} are sketched as follows.
\begin{enumerate}
	\item \textit{Initialization}.
	With the key $\textit{Seed}$ and logistic-sine map, generate a permutation vector  $\textbf{\textit{W}}$ and a substitution mask matrix $\textbf{\textit{K}}$.
	
	\item \textit{Random pixel insertion}.
	Generate $2\times M+2\times N+4$ random pixels, and then paste them around the four sides of the input plaintext  $\textbf{\textit{M}}$.
	Hence, an enlarged image $\textbf{\textit{MI}}$ with size $(H+2)\times (W+2)$ is obtained.	
	
	\item \textit{Permutation}.
	Shuffle the enlarged image $\textbf{\textit{MI}}$ using $\textbf{\textit{W}}$ and a certain row/column swapping approach.
	Analogously, the permutation procedure is generalized as 
	\[
	\textbf{\textit{P}}= \cal W(\textbf{\textit{MI}}).
	\]
	
	\item \textit{Substitution}.
	Stretch $\textbf{\textit{P}}$ column-by-column, and then perform pixel substitution according to Eq. (\ref{eq:Huadiffusion}).
	As indicated, a previous cipher pixel is linked.
	\begin{equation}    	
	c(i)=\\
	\left\{{\begin{aligned}{}
		&{p(i)\dot{+}k(i)\dot{+}p(L)} && i=1\\
		&{p(i)\dot{+}k(i)\dot{+}c(i-1)}&& i\in [2,L] \\
		\end{aligned}}. \right.
	\label{eq:Huadiffusion}
	\end{equation}	
	\item \textit{Iteration}. 
	The permutation-substitution network are iterated 2 times with independent $\textbf{\textit{W}}$ and $\textbf{\textit{K}}$.
\end{enumerate}

By adding random edge pixels, Hua's encryption scheme firstly enlarges the $H\times W$ input plaintext $\textbf{\textit{M}}$ into an intermediate image $\textbf{\textit{MI}}$ with size $(H+2)\times (W+2)$.
It is subsequently encrypted using two permutation-substitution iterations, hence, the ciphertext is also with size $(H+2)\times (W+2)$.
On the contrary, the input of Hua's decryption is a $(H+2)\times (W+2)$ ciphertext which will be decoded to an $H\times W$ image. 
Without loss of generality, we denote the random inserted variables as $\textbf{\textit{R}}$, and use symbol $||$  to demonstrate the linking operation of two matrices.
That means
\begin{equation}
\label{eq:MI_paste_MR}
\textbf{\textit{MI}} = \textbf{\textit{M}} || \textbf{\textit{R}}.
\end{equation}
In other words, $\textbf{\textit{MI}}$ is the input of the permutation-substitution network in the encryption, and it also denotes the decryption result before removing the edge pixels.
It should be emphasized that $\textbf{\textit{R}}$ is random and unobtainable in the encryption process, however, it is definite in the decryption though it is also unknown.
So is $\textbf{\textit{MI}}$.

As the sizes of $\textbf{\textit{M}}$ and $\textbf{\textit{C}}$ are not the same, it is difficult to obtain ${\cal {H}}_{(Hua)}(\Delta \textbf{\textit{M}})$ directly.
As a replacement, we build the relationship between the differential of $\textbf{\textit{MI}}$ with that of the ciphertext $\textbf{\textit{C}}$.
It is distinctively denoted as $\Delta \textbf{\textit{C}} = {\cal {H}}'_{(Hua)}(\Delta \textbf{\textit{MI}})$, without loss of generality. 
Firstly, we can easily rewrite Hua's substitution formula (Eq. (\ref{eq:Huadiffusion})) as  
\begin{equation} 
c(i)=p(L)\dot{+} {\dot \sum} _{j = 1}^i[p(j)\dot{+} k(j)],
\label{eq:Huadiffusion1}   	 	
\end{equation}
where $\textbf{\textit{P}}=\cal W(\textbf{\textit{MI}})$, i.e., the permutation ciphertext of the enlarged image. 
Referring to the deduction of ${\cal H}_{(basic)}(\Delta \textbf{\textit{M}})$ in Section \ref{sec:cryptanalysis_basicmodel} and taking Eqs. (\ref{eq:pv_permutation_1D}) and (\ref{eq:Huadiffusion1}) into consideration, we can get ${\cal {H}}'_{(Hua)}(\Delta \textbf{\textit{MI}})$ as
\begin{equation}
\left\{{\begin{aligned}{}
	&\Delta \textbf{\textit{C}}= {\cal {H}}'_{(Hua)}(\Delta \textbf{\textit{MI}})\\
	&\Delta c(i)=\Delta mi(w(L))\dot{+} {\dot \sum} _{j = 1}^i[\Delta mi(w(j))\dot{+} k(j)]
	\end{aligned}} \right.
,\label{eq:hua_transfer}
\end{equation}
where $w(j)$ is the element of the permutation vector $\textbf{\textit{W}}$.
Similar to  ${\cal H}_{(Zhou)}(\Delta \textbf{\textit{M}})$,  ${\cal H}'_{(Hua)}(\Delta \textbf{\textit{MI}})$ also consists of a series of permutation, modular addition and multiplication operations, therefore it has BAM properties.

Benefiting from the BAM properties of ${\cal H}'_{(Hua)}(\Delta \textbf{\textit{MI}})$, PCCA is found feasible for cracking Hua's cipher.
As a specified case, $LL=(H+2)\times (W+2)$ is defined in this subsection.
It refers to the pixel counts of the ciphertexts.
Complying with the PCCA's steps described in Section~\ref{sec:chosen_ciphertext_attack}, the input plaintext is recovered as follows.
\begin{enumerate}
	\item Referring to the first step of the attack, $LL+1$ chosen-ciphertexts, i.e., $\textbf{\textit{C}}_0, \cdots, \textbf{\textit{C}}_{LL}$, are firstly constructed, all of them are with size $(H+2)\times (W+2)$.
	
	\item Subsequently, $LL+1$ decryption results, $\textbf{\textit{M}}_0, \cdots, \textbf{\textit{M}}_{LL}$, are also obtainable.
	Yet, their sizes are all $H\times W$.
	\begin{itemize}
		\item As intermediate products, $\textbf{\textit{MI}}_0, \cdots, \textbf{\textit{MI}}_{LL}$ and $\textbf{\textit{R}}_0, \cdots, \textbf{\textit{R}}_{LL}$ exist during the decryption process.
		They are unknown, but not random.				
		
		\item According to Eq. (\ref{eq:MI_paste_MR}), $\textbf{\textit{MI}}_i = \textbf{\textit{M}}_i||\textbf{\textit{R}}_i$.		
		
	\end{itemize}	
	
	\item Further, we can calculate the differentials of the plaintexts according to $\Delta \textbf{\textit{M}}_{i}=\textbf{\textit{M}}_{i} \dot{-} \textbf{\textit{M}}_{0}$, $i\in [1, LL]$. 

	\begin{itemize}	
		
		\item Referring to Eq. (\ref{eq:MI_paste_MR}), we can get $\Delta \textbf{\textit{MI}}_{i} = \textbf{\textit{MI}}_{i} \dot{-} \textbf{\textit{MI}}_{0}=(\textbf{\textit{M}}_{i}||\textbf{\textit{R}}_{i}) \dot{-} (\textbf{\textit{M}}_0||\textbf{\textit{R}}_0)$. 
		
		\item As $||$ is matrices-linking operation and considering that pixels in different positions cannot affect each other in the modular subtraction of two matrices, thus $(\textbf{\textit{M}}_{i}||\textbf{\textit{R}}_{i}) \dot{-} (\textbf{\textit{M}}_0||\textbf{\textit{R}}_0)$=$(\textbf{\textit{M}}_{i}\dot{-} \textbf{\textit{M}}_0) || (\textbf{\textit{R}}_{i} \dot{-} \textbf{\textit{R}}_0)=\Delta \textbf{\textit{M}}_{i}||\Delta \textbf{\textit{R}}_{i}$. 
		
		\item To be concluded, $\Delta \textbf{\textit{MI}}_{i}=\Delta \textbf{\textit{M}}_{i}||\Delta \textbf{\textit{R}}_{i}$.
		
		\item Similarly, $\Delta \textbf{\textit{MI}}_{i}$ and $\Delta \textbf{\textit{R}}_{i}$ are unknown but not random, they are introduced for theoretical analysis.
		
	\end{itemize}
	
	\item For any ciphertext $\textbf{\textit{C}}=\{c(i),i\in[1,LL]\}$, assumed its plaintext as ${\textbf{\textit{M}}}$ whose differential between ${\textbf{\textit{M}}_0}$ is denoted as $\Delta {\textbf{\textit{M}}}$, we can get $\Delta \textbf{\textit{M}}$ according to Eq. (\ref{eq:calculation_delM_Hua}).
	\begin{equation}
	\label{eq:calculation_delM_Hua}
	\Delta \textbf{\textit{M}} = {\dot \sum} _{i = 1}^{LL}[c(i)\dot{\times}\Delta \textbf{\textit{M}}_i].
	\end{equation}
	\begin{itemize}		
		\item As  $\textbf{\textit{C}}_0$ is a zero image, we can therefore get $\Delta \textbf{\textit{C}}_i = \textbf{\textit{C}}_i \dot{-} \textbf{\textit{C}}_0 = \textbf{\textit{C}}_i$.
		
		\item Note that $\Delta \textbf{\textit{C}}_{i}$ and $\Delta \textbf{\textit{M}}_{i}$ have different sizes.
		
		\item Considering that ${\cal H}'_{(Hua)}(\Delta \textbf{\textit{MI}})$ has BAM properties, thus, 
		$\Delta \textbf{\textit{MI}} = {\dot \sum} _{i = 1}^{LL}[c(i)\dot{\times}\Delta \textbf{\textit{MI}}_i]$.
		
		\item Once again, because $||$ denotes matrices-liking operation and pixels in different positions cannot affect each other in modular addition/multiplication of two matrices, thus $c(i)\dot{\times}\Delta \textbf{\textit{MI}}_i=c(i)\dot{\times} (\Delta \textbf{\textit{M}}_{i}||\Delta \textbf{\textit{R}}_{i})=(c(i)\dot{\times} \Delta \textbf{\textit{M}}_{i})||(c(i)\dot{\times} \Delta \textbf{\textit{R}}_{i})$.	
		
		\item Furthermore, ${\dot \sum} _{i = 1}^{LL}[(c(i)\dot{\times} \Delta \textbf{\textit{M}}_{i})||(c(i)\dot{\times} \Delta \textbf{\textit{R}}_{i})]={\dot \sum} _{i = 1}^{LL}[(c(i)\dot{\times} \Delta \textbf{\textit{M}}_{i})]||{\dot \sum} _{i = 1}^{LL}[(c(i)\dot{\times} \Delta \textbf{\textit{R}}_{i})]$.
		
		\item Therefore, $\Delta \textbf{\textit{MI}}$ can be finalized as $\Delta \textbf{\textit{MI}}={\dot \sum} _{i = 1}^{LL}[(c(i)\dot{\times} \Delta \textbf{\textit{M}}_{i})]||{\dot \sum} _{i = 1}^{LL}[(c(i)\dot{\times} \Delta \textbf{\textit{R}}_{i})]$.

		\item Besides, it is known that $\Delta \textbf{\textit{MI}}=\Delta \textbf{\textit{M}}||\Delta \textbf{\textit{R}}$.
		
		\item Till now, we can come to Eq. (\ref{eq:calculation_delM_Hua}).		
	\end{itemize}
	
	\item Therefore, the plaintext is recovered as ${\textbf{\textit{M}}}=\Delta \textbf{\textit{M}} \dot{+} \textbf{\textit{M}}_0$.
\end{enumerate}

\subsection{Applicability to other ciphers
\cite{LAN2018133,borujeni2009chaotic,HUA2018148,HUA201580,hua2018cosine,hua2019imageAccess,hua2017filtering}}
\label{sec:cryptan_peers}

Except for the aforementioned case studies, image encryption schemes in \cite{LAN2018133,borujeni2009chaotic,HUA2018148,HUA201580,hua2018cosine,hua2019imageAccess,hua2017filtering}
share similar architecture, and also suffer from the reported cryptanalysis.

\begin{enumerate}
	\item \textit{Applicability to the cipher in \cite{LAN2018133}}.
	Two integrated chaotic systems are investigated for key schedule, and a new permutation approach is developed. 
	The permutation vector and substitution masks are fully depending on the secret key. 
	A slight difference is that the substitution is performed using modular subtraction, specifically, $c(i)=k(i)\dot{-} m(i)$.
	Referring to the cryptanalysis of the basic cipher given in Section \ref{sec:cryptanalysis_basicmodel}, it is easy to get the BAM properties of this cipher's DTF.
	Thus, PCCA is feasible.
	
	\item \textit{Applicability to the cipher in \cite{borujeni2009chaotic}}.
	Image permutation is performed in pixel-level, and the substitution is conducted with Eq. (\ref{eq:modaddiffusion}).
	In this cipher, logistic map, Tompkins-Paige algorithm and Tent map are employed for generating the permutation vector and substitution masks.
	These encryption elements are independent from the plaintexts. 
	The cryptanalysis of the basic cipher described in Section \ref{sec:cryptanalysis_basicmodel} is straightforward valid for this cipher.
	
	\item \textit{Applicability to the cipher in \cite{HUA2018148}}.
	This encryption scheme consists of four permutation-substitution iterations.
	The permutation is performed in pixel-level, while two previous ciphertexts are linked into the substitution as given in Eq. (\ref{eq:modaddc2diffusion}).
	Referring to the cryptanalysis of Zhou's encryption scheme described in Section~\ref{sec:zhoucryptanalysis}, this cipher is vulnerable to PCCA accordingly.
	The adopted novel permutation technique, complex logistic-sine-coupling map and more encryption rounds cannot promote the resistance against PCCA.
	
	\item \textit{Applicability to the cipher in \cite{HUA201580}}.
	This cipher employs sine-logistic modulation map for key schedule, and introduces Eq.~(\ref{eq:modaddc1diffusion}) for pixel substitution.
	A permutation, row-by-row and column-by-column substitutions jointly constitute the encryption core, which is iterated two times.
	Essentially, the encryption loop can be regarded as a two-layers permutation-substitution, where the second permutation is a $90^{\circ}$ clock rotation.
	It is easy to conclude that the DTF of this cipher is similar to Eq. (\ref{eq:hua_transfer}) by replacing $\Delta \textbf{\textit{MI}}$ as $\Delta \textbf{\textit{M}}$ \footnote{As aforementioned, cryptanalysis of Hua's cipher \cite{HUA2018134} in Section \ref{sec:huacryptanalysis} is given with the specific purpose to illustrate that inserting random pixels during the encryption process cannot promote the security against PCCA.}, also, it has BAM properties.
	As a consequence, PCCA is applicable. 
	
	\item \textit{Applicability to the cipher in \cite{hua2018cosine}}.
	Cosine-transform-based chaotic system is developed for key schedule, while Eq.~(\ref{eq:modaddc1diffusion}) is used for substitution.
	The encryption core, composed of a permutation, rotation and substitution, is repeated four times. 
	There are two features in this cipher.
	On one hand, there is an image rotation module between the permutation and substitution procedures. 
	Secondly, the substitution is performed secretly rather than the traditional sequential implementation. 
	By considering the secret-order substitution as a permutation-then-substitution (sequential) procedure, the encryption core turns to three permutation and one substitution procedures.
	Obviously, we can use one permutation vector to synthesize the three permutation modules, and finalize the iteration as a permutation-substitution network.
	Similar to that of \cite{HUA201580}, PCCA is applicable too.
	
	\item \textit{Applicability to the ciphers in \cite{hua2019imageAccess,hua2017filtering}.}
	Both ciphers outfit the iterative permutation-substitution structure, yet the substitution is derived from the image filtering concept. 
	Technically, many neighbor pixels are linked for substitution.
	Referring to the cryptanalysis when linking one or two adjacent pixels for substitution, as given in Sections \ref{sec:zhoucryptanalysis} and \ref{sec:huacryptanalysis}, it is reasonable to infer that the DTFs of both ciphers \cite{hua2019imageAccess,hua2017filtering} are composed of a series of permutation, modular addition and multiplication operations.
	They are similar to Eqs.~(\ref{eq:zhou_transfer}) and~(\ref{eq:hua_transfer}), yet may own different but definite coefficients.
	Accordingly, they also have BAM properties.
	As a consequence, the ciphers in \cite{hua2019imageAccess,hua2017filtering} own identical security drawback of peer schemes and are vulnerable against PCCA.
\end{enumerate}

\subsection{Numerical experiments}
\label{sec:experiments}
This subsection gives numerical experimental results and relative discussions.
The experiments are implemented in Matlab 2014a platform, and the source codes are open accessible at \url{https://github.com/lurenjia212/Break_modadd}.

Zhou's cipher \cite{YicongTC6940279} is firstly introduced for a step-by-step illustration.
Without loss of generality, the image size is set as 9 whose particles are 8-bit gray pixels, and the default key embedded in Zhou's source code is adopted \footnote{A default key is embedded in Zhou's source code, which is open accessible via \url{https://www.fst.um.edu.mo/en/staff/documents/fstycz/Hua2015TC.rar}.}.
Using this key, Alice encrypts an image 
\[\textbf{\textit{M}}=\{0, 15, 33, 47, 65, 165, 56, 96, 255\},\]
and get the ciphertext 
\[\textbf{\textit{C}}=\{29,67,  144,143,74,127,101,24,139\},\]
which is eavesdropped by Eve. 
Subsequently, Eve tries to recover the plaintext without the secret key. 
Completely complying with the attack procedures in Section \ref{sec:chosen_ciphertext_attack}, the signal recovery processes are illustrated as follows.
\begin{enumerate}
	\item Construct $9+1=10$ chosen-ciphertxts, they are denoted as $\textbf{\textit{C}}_0, \textbf{\textit{C}}_1, \cdots, \textbf{\textit{C}}_{9}$, respectively. 
	\[\begin{array}{*{20}{l}}
	\textbf{\textit{C}}_0=\{0,0,0,0,0,0,0,0,0\}  \\
	\textbf{\textit{C}}_1=\{1,0,0,0,0,0,0,0,0\}  \\
	\textbf{\textit{C}}_2=\{0,1,0,0,0,0,0,0,0\}  \\
	\textbf{\textit{C}}_3=\{0,0,1,0,0,0,0,0,0\}  \\
	\textbf{\textit{C}}_4=\{0,0,0,1,0,0,0,0,0\}  \\
	\textbf{\textit{C}}_5=\{0,0,0,0,1,0,0,0,0\}  \\
	\textbf{\textit{C}}_6=\{0,0,0,0,0,1,0,0,0\}  \\
	\textbf{\textit{C}}_7=\{0,0,0,0,0,0,1,0,0\}  \\
	\textbf{\textit{C}}_8=\{0,0,0,0,0,0,0,1,0\}  \\
	\textbf{\textit{C}}_9=\{0,0,0,0,0,0,0,0,1\}  \\
	\end{array}\]
	
	\item Under the assumption of chosen-ciphertext attack, their corresponding plaintexts are obtained as 
    \[\begin{array}{*{20}{l}}
    \textbf{\textit{M}}_0~=\{85,16,228,187,2,230,109,110,193\}  \\
    \textbf{\textit{M}}_1~=\{86,14,227,189,3,230,109,110,193\}  \\
    \textbf{\textit{M}}_2~=\{85,17,226,  186,    4,  231,  109,  110,  193\}  \\
    \textbf{\textit{M}}_3~=\{85,   16,  229,  185,    1,  232 , 110 , 110, 193\}  \\
    \textbf{\textit{M}}_4~=\{84,   16,  228,  188,    0,  229,  111 , 111,  193\}  \\
    \textbf{\textit{M}}_5~=\{82,   15,  228,  187,    3,  228 , 108 , 112,  194\}  \\
    \textbf{\textit{M}}_6~=\{85,   13,  227,  187,    2,  231 , 107 , 109 , 195\}  \\
    \textbf{\textit{M}}_7~=\{90,   16,  225,  186,    2,  230 , 110 , 108 , 192\}  \\
    \textbf{\textit{M}}_8~=\{86,   19,  227,  186,    2,  230 , 109 , 111 , 191\}  \\
    \textbf{\textit{M}}_9~=\{83,   15,  230,  188,    2,  230 , 109 , 110 , 194\}  \\
    \end{array}.\]
	
	\item  The differentials of the plaintexts are  
	\[\begin{array}{*{20}{l}}
	\Delta \textbf{\textit{M}}_1=\textbf{\textit{M}}_1\dot{-}\textbf{\textit{M}}_0=\{1,  254,  255 ,   2,    1,    0,    0 ,   0 ,   0\}\\
	\Delta \textbf{\textit{M}}_2=\textbf{\textit{M}}_2\dot{-}\textbf{\textit{M}}_0=\{0,    1,  254,  255,   2,    1,    0 ,   0 ,   0\}\\
	\Delta \textbf{\textit{M}}_3=\textbf{\textit{M}}_3\dot{-}\textbf{\textit{M}}_0=\{0,    0,    1,  254,  255,    2,    1 ,   0 ,   0 \}\\
	\Delta \textbf{\textit{M}}_4=\textbf{\textit{M}}_4\dot{-}\textbf{\textit{M}}_0=\{255 ,   0,    0,    1,  254,  255,    2,    1 ,   0 \}\\
	\Delta \textbf{\textit{M}}_5=\textbf{\textit{M}}_5\dot{-}\textbf{\textit{M}}_0=\{253,  255,    0,    0,    1,  254,  255 ,   2,    1\}\\
	\Delta \textbf{\textit{M}}_6=\textbf{\textit{M}}_6\dot{-}\textbf{\textit{M}}_0=\{0 , 253,  255,    0 ,   0,    1,  254,  255,    2 \}\\
	\Delta \textbf{\textit{M}}_7=\textbf{\textit{M}}_7\dot{-}\textbf{\textit{M}}_0=\{5,    0,  253,  255 ,   0,    0,    1,  254 , 255 \}\\
	\Delta \textbf{\textit{M}}_8=\textbf{\textit{M}}_8\dot{-}\textbf{\textit{M}}_0=\{ 1 ,   3,  255,  255,   0,    0,    0 ,   1,  254 \}\\
	\Delta \textbf{\textit{M}}_9=\textbf{\textit{M}}_9\dot{-}\textbf{\textit{M}}_0=\{254 , 255 ,   2,    1,   0,    0,    0,    0,    1 \}\\
	\end{array}.\]

	\item As to the eavesdropped ciphertext  $\textbf{\textit{C}} = \{29, 67, 144,$ $143,74,$$127,101,24,139\}$, assume its plaintext as ${\textbf{\textit{M}}}$ whose differential between ${\textbf{\textit{M}}_0}$ is further denoted as $\Delta \textbf{\textit{M}}$, Eve can get $\Delta \textbf{\textit{M}}$ through
	 \begin{equation*}
	{
		\begin{aligned}
		\Delta \textbf{\textit{M}} & = {\dot \sum} _{i = 1}^9[c(i)\dot{\times}\Delta \textbf{\textit{M}}_i]  \\
		&=\{ 171,  255,61,  116,   63,  191,  203,  242,   62\}\\
		\end{aligned}.
	}
	\end{equation*}
	
	\item Finally, Eve recover the plaintext $\textbf{\textit{M}}$ according to 
	\begin{equation*}
	{
		\begin{aligned}
		{\textbf{\textit{M}}}&=\Delta \textbf{\textit{M}}\dot{+} \textbf{\textit{M}}_0  \\
		&=\{ 171,  255,61,  116,   63,  191,  203,  242,   62\}\\
		&~\dot{+}\{85,16,228,187,2,230,109,110,193\}\\
		&=\{0, 15, 33, 47, 65, 165, 56, 96, 255\}\\
		\end{aligned}.
	}
	\end{equation*}
\end{enumerate}
As can be observed, the recovered plaintext is exactly the same with Alice's original input $\textbf{\textit{M}}$.

\begin{figure}
	\centering
	\subfigure[]{\includegraphics[width=2.1cm]{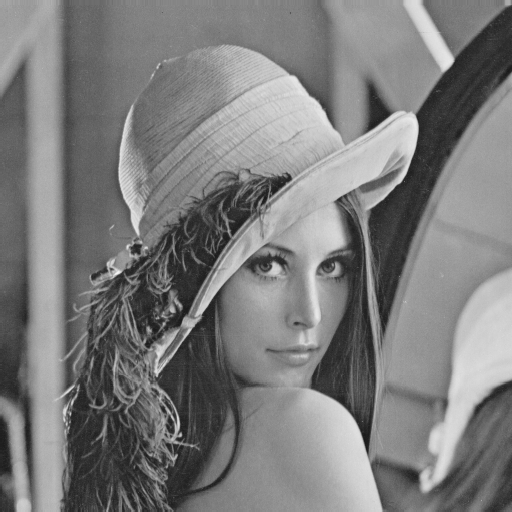}}
	\subfigure[]{\includegraphics[width=2.1cm]{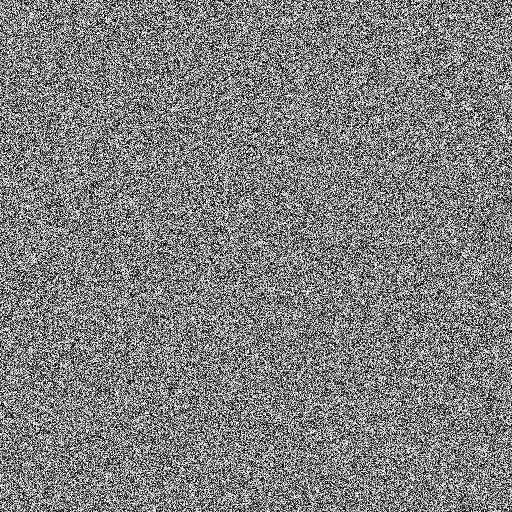}}
	\subfigure[]{\includegraphics[width=2.1cm]{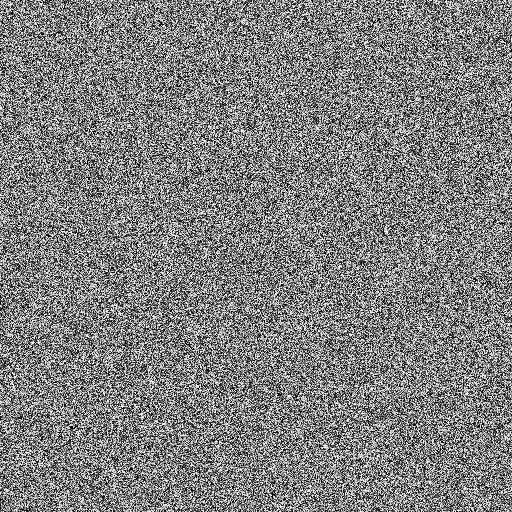}}
	\subfigure[]{\includegraphics[width=2.1cm]{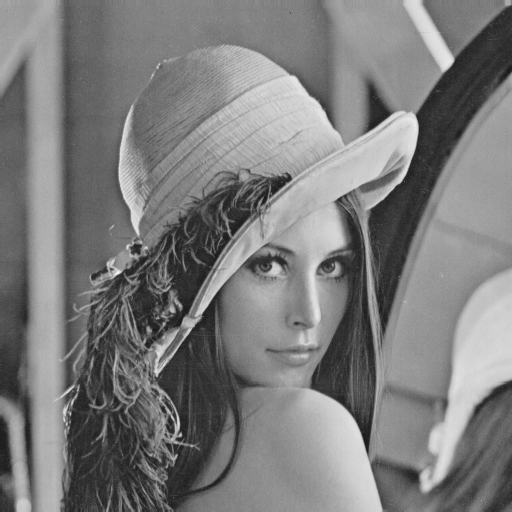}}	
	\subfigure[]{\includegraphics[width=2.1cm]{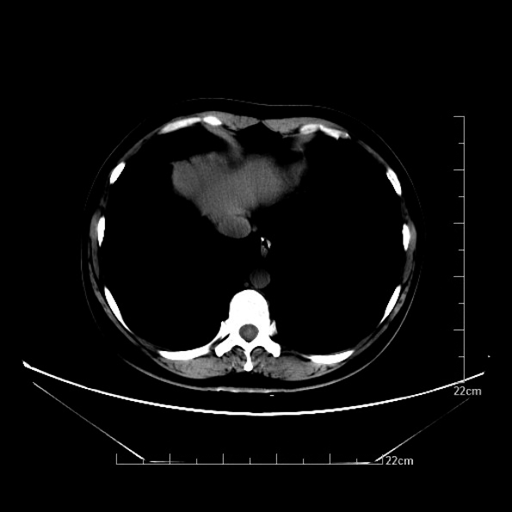}}
	\subfigure[]{\includegraphics[width=2.1cm]{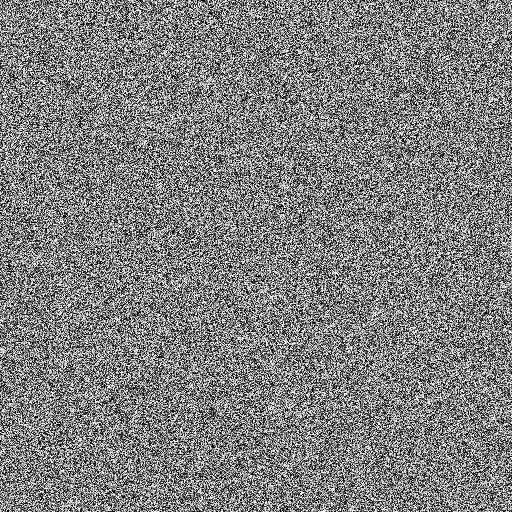}}
	\subfigure[]{\includegraphics[width=2.1cm]{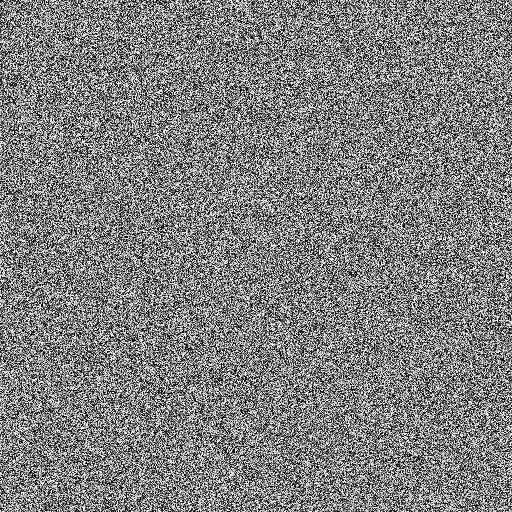}}
	\subfigure[]{\includegraphics[width=2.1cm]{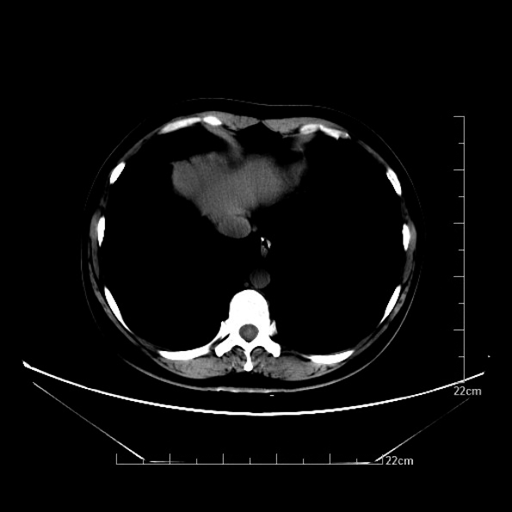}}
	\subfigure[]{\includegraphics[width=2.1cm]{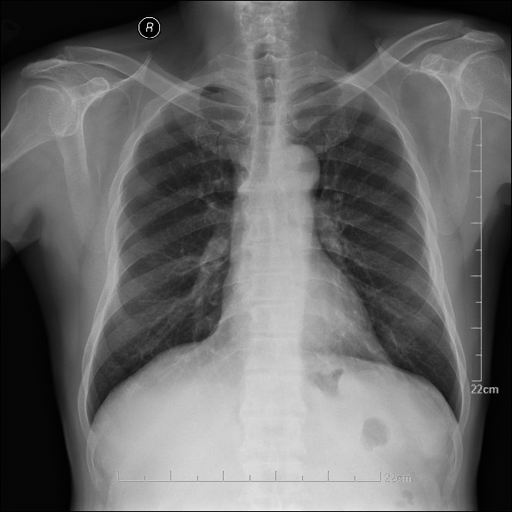}}
	\subfigure[]{\includegraphics[width=2.1cm]{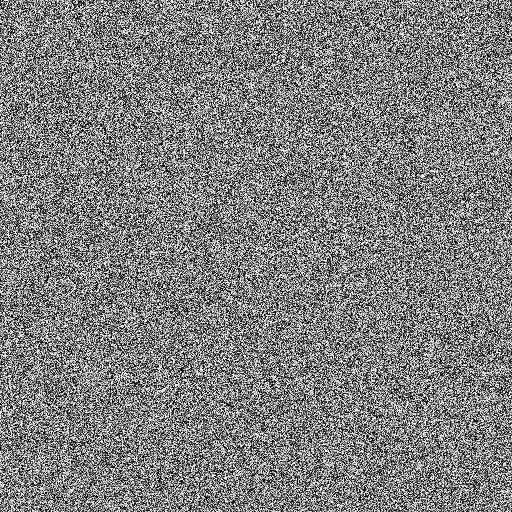}}
	\subfigure[]{\includegraphics[width=2.1cm]{Hua_del_m.png}}
	\subfigure[]{\includegraphics[width=2.1cm]{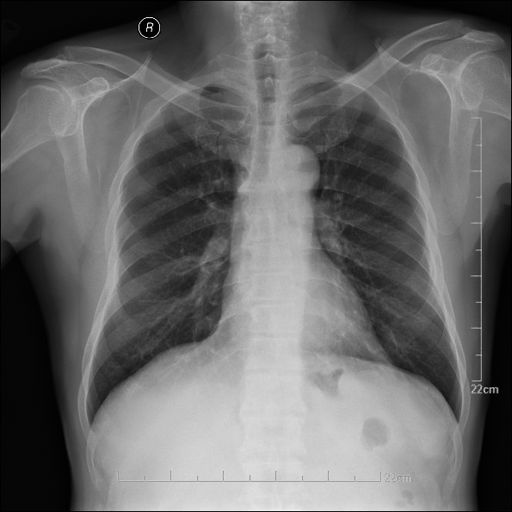}}
	\caption{Experiment results of the PCCA: (a)-(d), (e)-(h), and (i)-(l) demonstrate the plaintext, ciphertext, calculated differential between plaintext and $\textbf{\textit{M}}_0$, the recovered image of Lan's \cite{LAN2018133}, Zhou's \cite{YicongTC6940279} and Hua's \cite{HUA2018134} ciphers, respectively.}
	\label{fig:experments}
\end{figure}

Besides, a gray-scale Lena, 16-bit precision X-ray and CT images are introduced into Lan's \cite{LAN2018133}, Zhou's \cite{YicongTC6940279} and Hua's \cite{HUA2018134} ciphers, respectively. 
With PCCA, the plaintexts have been successfully recovered, as demonstrated in Fig.~\ref{fig:experments} where corresponding results of Lan's, Zhou's and Hua's ciphers are listed in the $1^{st}$ to $3^{rd}$ row, respectively.
The results are reproducible, interested readers are encouraged to refer to the source codes for details and extensions. 

\section{Conclusions}
\label{sec:conclusions}
In this paper, we have studied the security of a family of image encryption schemes.
The substitution of the involved ciphers is implemented by a modular addition function, and the encryption elements in both permutation and substitution procedures are generated by the employed nonlinear dynamics. 
We found the linear relationship between the differential of plaintexts with that of ciphertexts.
We then proposed a universal chosen-ciphertext attack which can decrypt the ciphertext without retrieving the secret key or equivalent encryption elements. 
Taking certain representative ciphers as case study, the proposed attack has been experimentally validated. 
It was also illustrated that the routine improvements, including new dynamic phenomena and permutation techniques, are not able to remedy the reported security flaws.
We hope the proposed cryptanalysis will be beneficial for the theoretical design and security evaluation of future image ciphers. 

\begin{appendix}[Proof of Property \ref{property:default_transitivity}]
	
For image ciphers with iterative architecture, the output in $(i-1)^{th}$ layer will be the input of the next encryption round, as shown in Eq. (\ref{eq:i_round_ciphertext}).
Therefore, we can get
\begin{equation}
\begin{aligned}
\Delta \textbf {\textit{C}}^{(i)}&={\cal H}^{(i)}_{(basic)}(\Delta \textbf{\textit{M}}^{(i)}) \\
&={\cal H}^{(i)}_{(basic)}(\Delta \textbf {\textit{C}}^{(i-1)})\\
&={\cal H}^{(i)}_{(basic)}[{\cal H}^{(i-1)}_{(basic)}(\Delta \textbf {\textit{C}}^{(i-2)})]\\
&=\dots\\
&={\cal H}^{(i)}_{(basic)}\{{\cal H}^{(i-1)}_{(basic)}[\dots {\cal H}^{(1)}_{(basic)}(\Delta \textbf {\textit{C}}^{(0)})]\}\\
&={\cal H}^{(i)}_{(basic)}\{{\cal H}^{(i-1)}_{(basic)}[\dots {\cal H}^{(1)}_{(basic)}(\Delta \textbf {\textit{M}}^{(1)})]\}
\end{aligned}.
\end{equation}
In other words, 
\begin{equation*}	
\begin{aligned}
\label{eq:default_hndefinition1n2}
{\cal H}^{(1)-(N)}_{(basic)}(\Delta \textbf{\textit{M}}^{(1)})	={\cal H}^{(N)}_{(basic)}\{{\cal H}^{(N-1)}_{(basic)}[\dots {\cal H}^{(1)}_{(basic)}(\Delta \textbf{\textit{M}}^{(1)})]\}.
\end{aligned}
\end{equation*}
As ${\cal H}^{(i)}_{(basic)}(\Delta \textbf{\textit{M}}), i\in[1,N]$ is bijective, modular additive and multipliable, ${\cal H}^{(1)-(N)}_{(basic)}(\Delta \textbf{\textit{M}}^{(1)})$ consequently has the properties of bijectivity, additivity and multiplicability.

Hence completes the proof.	
\end{appendix}

\bibliographystyle{IEEEtran}
\bibliography{reference_modadd_R2}

\end{document}